\documentclass[12pt]{iopart}
\pdfoutput=1 
\usepackage[pdftex]{graphicx,color}
\usepackage{amssymb}
\usepackage{epsfig}
\usepackage{subfigure}
\usepackage{textcomp}
\usepackage{cite}
\usepackage{float}
\usepackage{bm}% bold math
\expandafter\let\csname equation*\endcsname\relax
\expandafter\let\csname endequation*\endcsname\relax
\usepackage{amsmath}
\graphicspath{ {./figs/} }
\def \equi#1{\mathrel{\mathop{\kern 0pt\sim}\limits_{#1}}}

\begin{document}
\title{The Advantage of Foraging Myopically}

\author{C. L. Rager} \address{\"Ojendorfer H\"ohe 56B, 22117 Hamburg, Germany}
\address{Santa Fe Institute, 1399 Hyde Park Road, Santa
  Fe, New Mexico 87501, USA}

\author{U. Bhat} \address{School of Natural Sciences, University of
  California Merced, Merced, California 95343, USA}

\author{O. B\'enichou} \address{Laboratoire de Physique Th\'eorique de la
  Mati\`ere Condens\'ee (UMR CNRS 7600), Universit\'e Pierre et Marie Curie,
  4 Place Jussieu, 75252 Paris Cedex France}

\author{S. Redner} \address{Santa Fe Institute, 1399 Hyde Park Road, Santa
  Fe, New Mexico 87501, USA}

\begin{abstract}
  We study the dynamics of a \emph{myopic} forager that randomly wanders on a
  lattice in which each site contains one unit of food.  Upon encountering a
  food-containing site, the forager eats all the food at this site with
  probability $p<1$; otherwise, the food is left undisturbed.  When the
  forager eats, it can wander $\mathcal{S}$ additional steps without food
  before starving to death.  When the forager does not eat, either by not
  detecting food on a full site or by encountering an empty site, the forager
  goes hungry and comes one time unit closer to starvation.  As the forager
  wanders, a multiply connected spatial region where food has been
  consumed---a desert---is created.  The forager lifetime depends
  non-monotonically on its degree of myopia $p$, and at the optimal myopia
  $p=p^*(\mathcal{S})$, the forager lives much longer than a normal forager
  that always eats when it encounters food.  This optimal lifetime grows as
  $\mathcal{S}^2/\ln\mathcal{S}$ in one dimension and faster than a power law
  in $\mathcal{S}$ in two and higher dimensions.

\end{abstract}
% \maketitle

\section{Introduction and Model}

\begin{figure}[ht]
  \centerline{\subfigure[]{\includegraphics[width=0.34\textwidth]{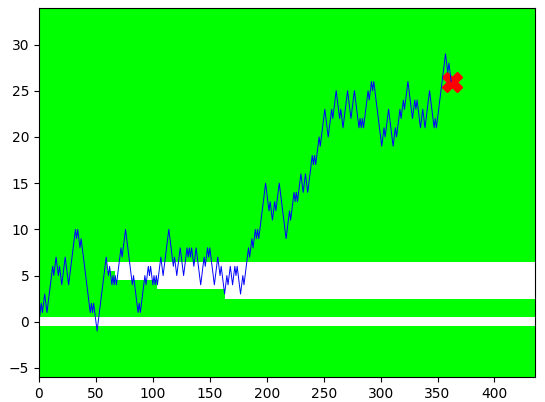}}
\subfigure[]{\includegraphics[width=0.34\textwidth]{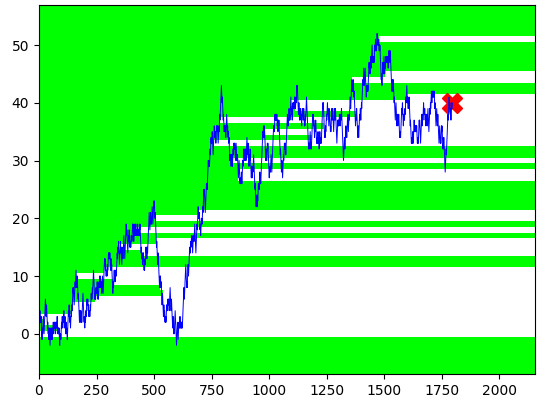}}
\subfigure[]{\includegraphics[width=0.34\textwidth]{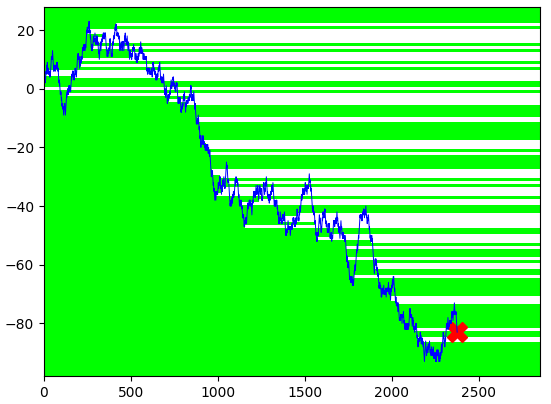}}}
\centerline{\subfigure[]{\includegraphics[width=0.5\textwidth]{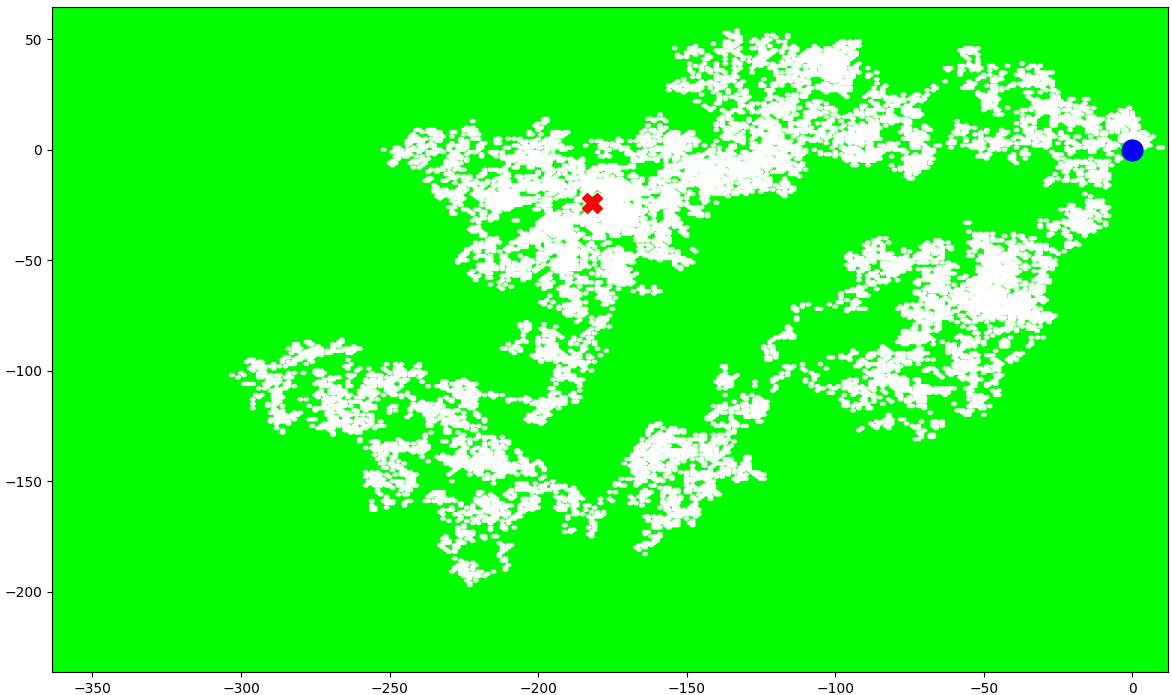}}}
\caption{Typical space-time trajectories of a myopic forager with
  $\mathcal{S}=200$ and $p=p^*\approx 0.035$ in dimension $d=1$ (a)--(c) and
  $\mathcal{S}=100$ and $p=0.2$ in $d=2$ (d).  Green shaded area represents
  food and white space denotes desert.  In $d=2$, the starting location of
  the forager is indicated by the blue dot. }
  \label{space-time}
\end{figure}

In this work, we extend of the starving random walk model of
foraging~\cite{BR14,CBR16} to the situation where the forager is myopic.
Whenever such a forager comes to a site that contains food, all the food at
this site is eaten with probability $p$, while the food is left undisturbed
with probability $1-p$.  In the limiting case of $p=1$, the forager always
consumes food when it is encountered.  This rule corresponds to the original
starving random walk, which here we term the normal forager.  We want to
understand the role of myopia---quantified by $p$---on the foraging dynamics
and on the geometry of the ``desert'', the region where food has been
consumed.  Our main results are: (a) the forager lifetime depends
non-monotonically on $p$, (b) at an optimal value of $p$, the forager
lifetime is much longer than that of a normal forager (with $p=1$), and (c)
the average geometry of the desert has a simple character, even though the
desert geometry for each individual trajectory is complex
(Fig.~\ref{space-time}).

Foraging is a fundamental biological process that has been extensively
investigated and documented in the ecology literature (see e.g.,
Refs.~\cite{MP66,C76,PPC77,SK86,OB90,B91}).  In classic theories of foraging,
a typical assumption is that the forager has complete knowledge of its
environment and makes rational decisions about when to continue exploiting a
local resource and when to explore a new search domain.  The starving random
walk model represents a complementary perspective in which the forager has no
knowledge of its environment and uses naive decision rules to search for
resources.

In the starving random walk model, a forager performs a random walk on an
infinite lattice, independent of the presence of absence of resources in its
local neighborhood~\cite{BR14,CBR16}.  When the forager lands on a
food-containing site, all the food at this site is consumed.  Upon eating, a
forager is fully satiated and can hop $\mathcal{S}$ additional steps without
again eating before it starves.  Upon landing on an empty site, the forager
goes hungry and comes one time unit closer to starvation.  The forager
starves when it last ate $\mathcal{S}$ steps in the past.  We may therefore
view $\mathcal{S}$ as the metabolic storage capacity of the forager.  Because
there is no resource replenishment, the forager is doomed to ultimately
starve and the basic question is: when does the forager starve?

This starving random walk model was recently extended to incorporate, in a
minimalist way, various aspects of real foraging.  As one
example~\cite{BRB17a,BRB17b}, a forager was endowed with the attribute of
greed, in which it moves preferentially towards food if the forager has a
choice between hopping to an empty site or a food-containing site in its
local neighborhood.  It was found that there exists an optimal greediness
that maximizes the forager lifetime in two dimensions, and an optimal value
of \emph{negative} greed (where the forager tends to avoid food in its local
neighborhood) that maximizes the lifetime in one dimension.  The starving
random walk was also extended to incorporate frugality, in which the forager
eats only if it is nutritionally depleted beyond a specified level when it
lands on a food-containing site~\cite{BBKR18}.  It was found that the forager
lifetime is maximized at an optimal frugality level.

The issues that we address in this work are: How does the myopia of a forager
affect its lifetime and the geometry of the desert that is created?  An
important consequence of myopia is that the desert is no longer simply
connected (Fig.~\ref{space-time}).  In dimension $d=1$, the desert consists
of multiple empty segments that are interspersed with oases---food-containing
segments.  As the forager wanders, it may nucleate a new desert segment when
it eats food within a previously undepleted region; conversely, the forager
may consume all the food in an oasis thereby joining disconnected desert
segments.  The connectedness of the desert in the original starving random
walk model was a crucial feature that allowed for an asymptotic solution of
the lifetime in $d=1$~\cite{BR14,CBR16}.  The multiple connectedness of the
desert (Figs.~\ref{space-time}(a)--(c)) for the myopic forager introduces a
new layer of complexity to this challenging non-Markovian process; the
problem in $d>1$ is geometrically even more complex
(Fig.~\ref{space-time}(d)).

In the next section, we first present a heuristic argument that accounts for
the behavior of the forager lifetime for small $p$ in any spatial dimension.
We also argue that the lifetime must depend non-monotonically on the myopia
parameter $p$, at least in low spatial dimensions.  In the following two
sections, we present simulation results for the optimal myopia value and for
the forager lifetime at the optimal myopia in spatial dimensions $d=1$, 2 and
3. We find that this maximal forager lifetime grows as
$\mathcal{S}^2/\ln\mathcal{S}$ $d=1$, and grows faster than any power law in
$\mathcal{S}$ for $d\geq 2$.  In both $d=1$ and $d=2$, the average density
profile of the desert decays exponentially in the distance from the starting
point of the forager.

\section{Heuristics}
\label{sec:heu}

Because of the geometrical complication that the myopic forager carves out a
multiply-connected desert, the approach used to analyze the dynamics of the
normal forager in a single-segment desert, is not applicable
here~\cite{BR14,CBR16}.  However, we can understand the behavior of the
lifetime when $p$ is within a suitable range.  The extreme case of
$p\ll \mathcal{S}^{-1}$ is uninteresting because the forager typically does
not eat before it starves, so that its lifetime equals $\mathcal{S}$.  Thus
we examine the case where $p$ is small, but with $p> \mathcal{S}^{-1}$, so
that the forager typically eats multiple times before it starves.

\begin{figure}[ht]
  \center{\includegraphics[width=0.65\textwidth]{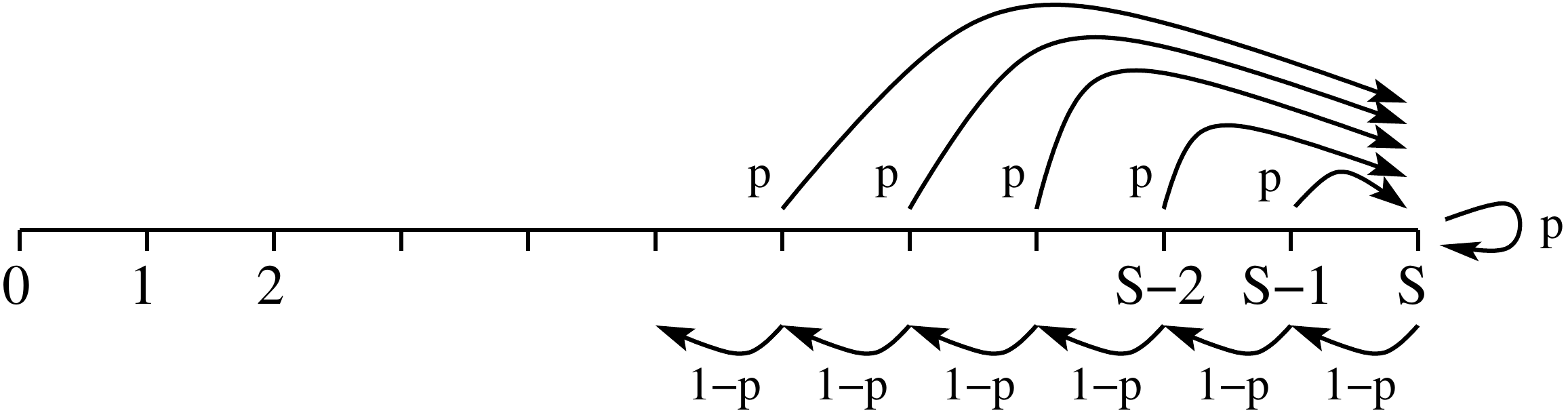}}
  \caption{State space of the myopic forager in the limit of small $p$.}
  \label{state-space}
\end{figure}

In this range of $p\gtrsim \mathcal{S}^{-1}$, but with $p\ll1$, it is
unlikely that the forager will revisit a site where food was previously
consumed.  Thus we assume that the forager always lands on a full site and
check the validity of this assumption at the end of the calculation.  If
there is no depletion, there are only two outcomes after the forager takes a
single step: it either eats with probability $p$, or does not eat with
probability $1-p$.  The state space for this process is depicted in
Fig.~\ref{state-space}; this same state-space geometry also arises in models
of kinetic proofreading~\cite{H74,MBN09,BMN10} and in the starving random
walk in the mean-field limit~\cite{CBR16}.  The forager starts in the fully
satiated state, corresponding to the right edge of the interval of length
$\mathcal{S}$ in the figure.  When the forager does not eat, it comes one
time unit closer to starvation and thus hops one step to the left in state
space.  When the forager eats, it is fully satiated and hops all the way to
the right edge of the interval.  Starvation corresponds to the forager
reaching site 0.  The forager lifetime $\mathcal{T}$ is just the mean time
for the particle in this state space to first reach site 0 when starting from
site $\mathcal{S}$ (Fig.~\ref{state-space}).  This time can be determined by
the formalism of Ref.~\cite{R01} and was previously computed in
Ref.~\cite{CBR16} to be:
\begin{equation}
  \label{T}
  \mathcal{T} =\frac{1}{p}\big[(1-p)^{-\mathcal{S}}-1\big]\,.
\end{equation}

\begin{figure}[ht]
  \center{\subfigure[]{\includegraphics[width=0.475\textwidth]{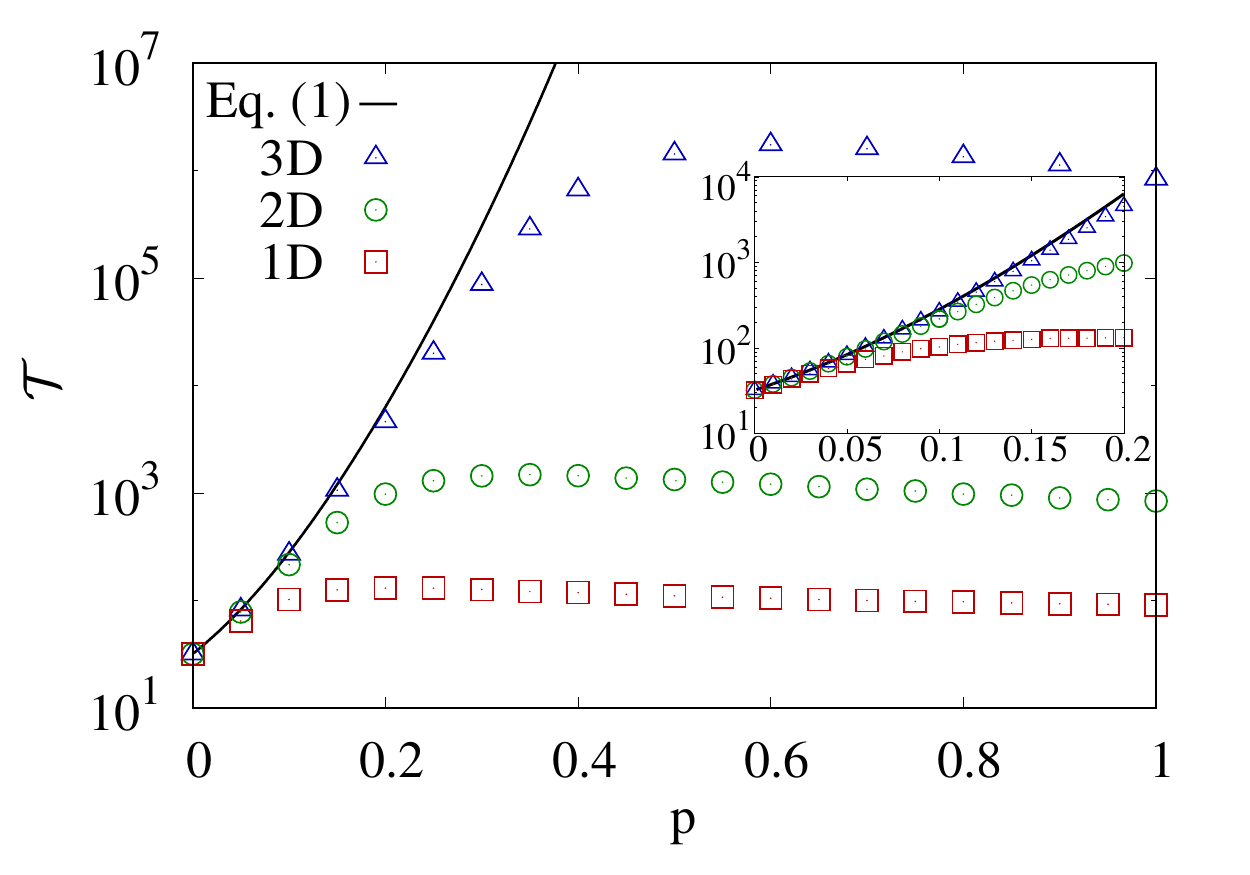}}\quad
    \subfigure[]{\includegraphics[width=0.45\textwidth]{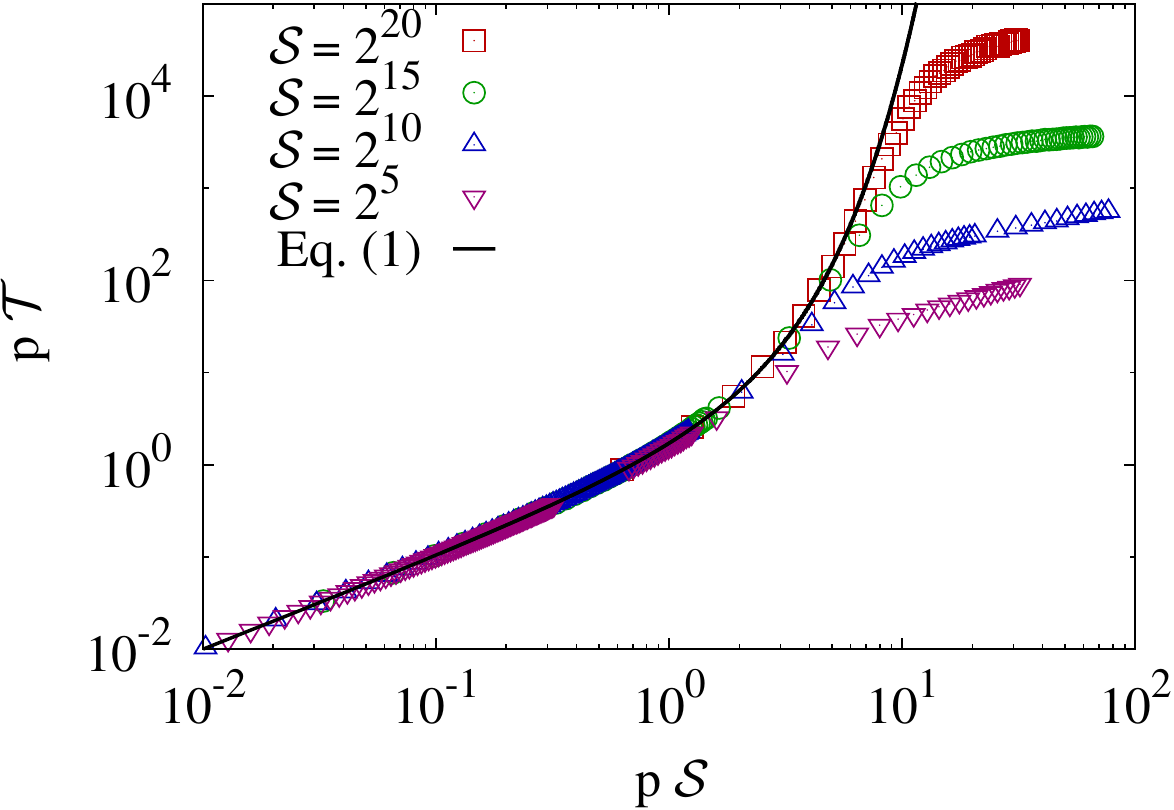}}}
  \caption{(a) Comparison of the heuristic lifetime from Eq.~\eqref{T} with
    simulation data in $d=1$, 2, and 3 for $\mathcal{S}=32$.  The inset shows
    detail for $p<0.2$.  (b) Comparison of Eq.~\eqref{T} expressed in scaled
    form with simulation data in $d=1$ for $\mathcal{S}$ up to $2^{20}$.}
  \label{compare}
\end{figure}

Equation~\eqref{T} should hold as long as the density of food eaten over the
spatial range where the forager wanders throughout its lifespan is small.
After $\mathcal{T}$ steps, this spatial range is of the order of
$\sqrt{\mathcal{T}}$ in dimension $d=1$.  Thus the density of food eaten
within this spatial range is of the order of
$p\mathcal{T}/\sqrt{\mathcal{T}}=p\sqrt{\mathcal{T}}$, which should be be
less than 1 for the assumption of no depletion to be valid.  We therefore
substitute the lifetime from \eqref{T} into $p\sqrt{\mathcal{T}}<1$ to give
$\ln(1-p)>(\ln p)/\mathcal{S}$.  Expanding the logarithm to lowest order
gives $p<-\ln p/\mathcal{S}$, or $p<(\ln \mathcal{S})/\mathcal{S}$.  In $d$
dimensions, the density of food eaten after $\mathcal{T}$ steps is given by
$p\mathcal{T}/\mathcal{T}^{d/2}$.  Requiring this density to be small gives
$p<1$ in $d=2$ and no constraint on $p$ for $d>2$.  These predictions accord
with simulation data for $\mathcal{S}=32$ (Fig.~\ref{compare}(a)), which is
the largest value of $\mathcal{S}$ that we can practically simulate in $d=3$,
and for $\mathcal{S}$ up to $2^{20}$ in $d=1$ (Fig.~\ref{compare}(b)).  The
agreement between the data and Eq.~\eqref{T} holds over a larger range of $p$
as the dimension increases, as follows from our argument.

\section{Simulations in One Dimension}

\begin{figure}[ht]
\center{\includegraphics[width=0.55\textwidth]{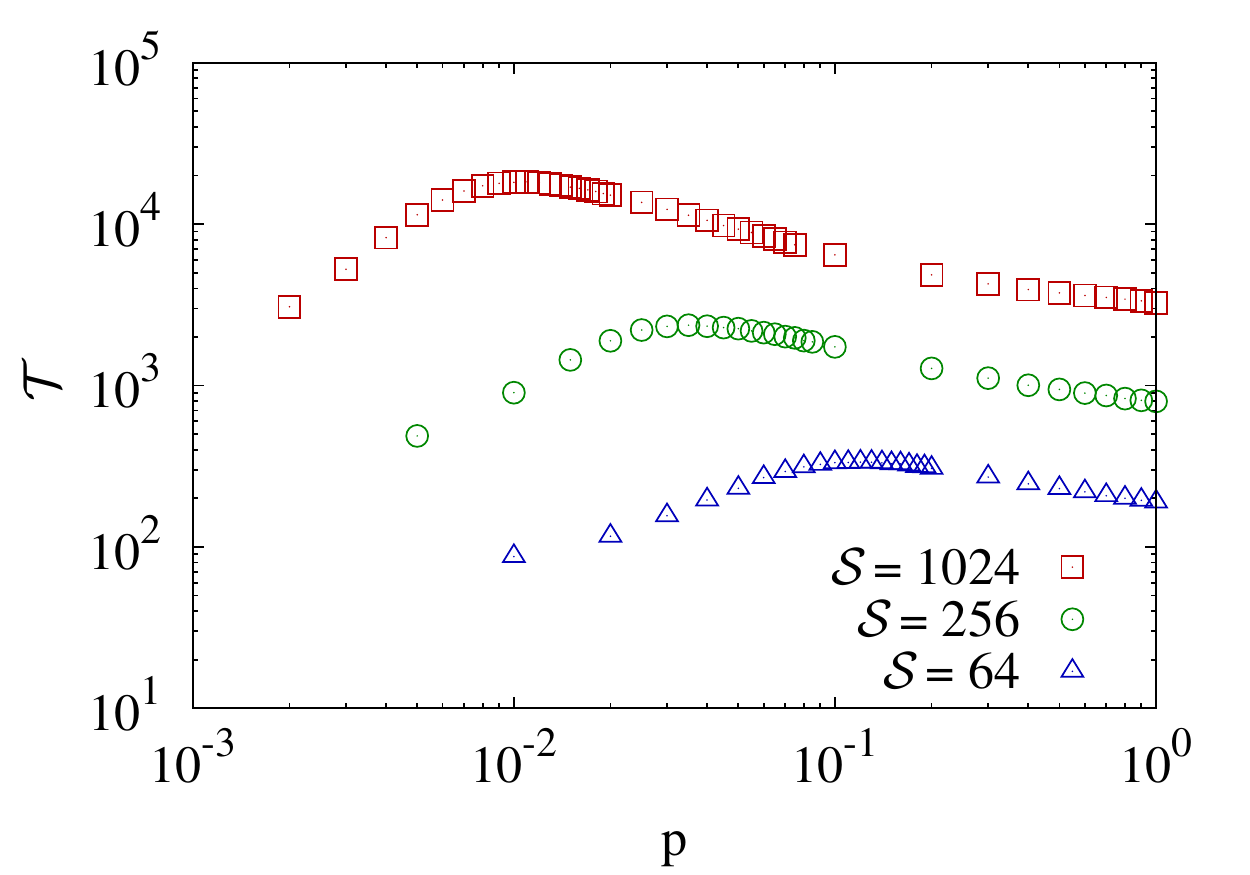}}
\caption{Forager lifetime versus $p$ in $d=1$ for three representative
  $\mathcal{S}$ values.}
  \label{t-vs-p-1d}
\end{figure}

\begin{figure}[ht]
\centerline{\subfigure[]{\includegraphics[width=0.45\textwidth]{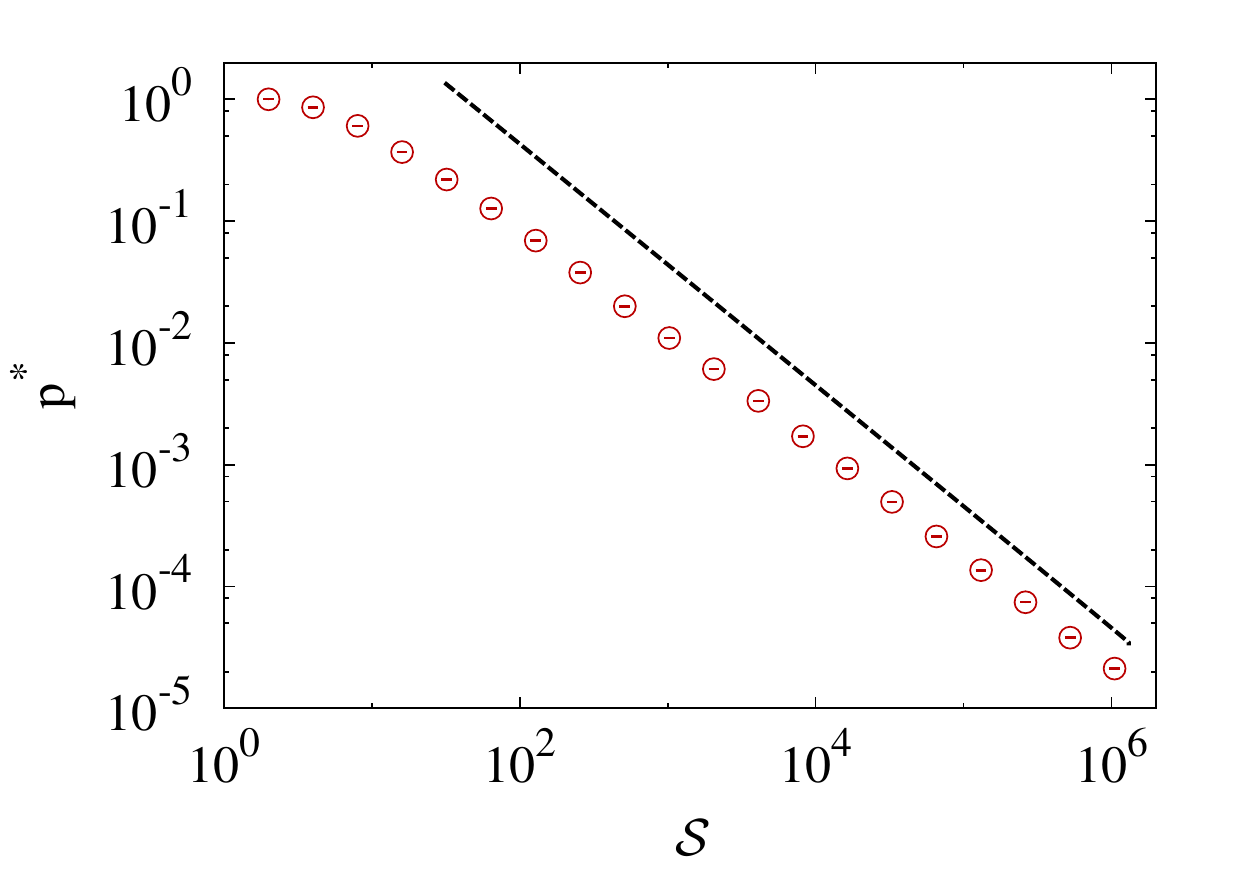}}
  \subfigure[]{\includegraphics[width=0.45\textwidth]{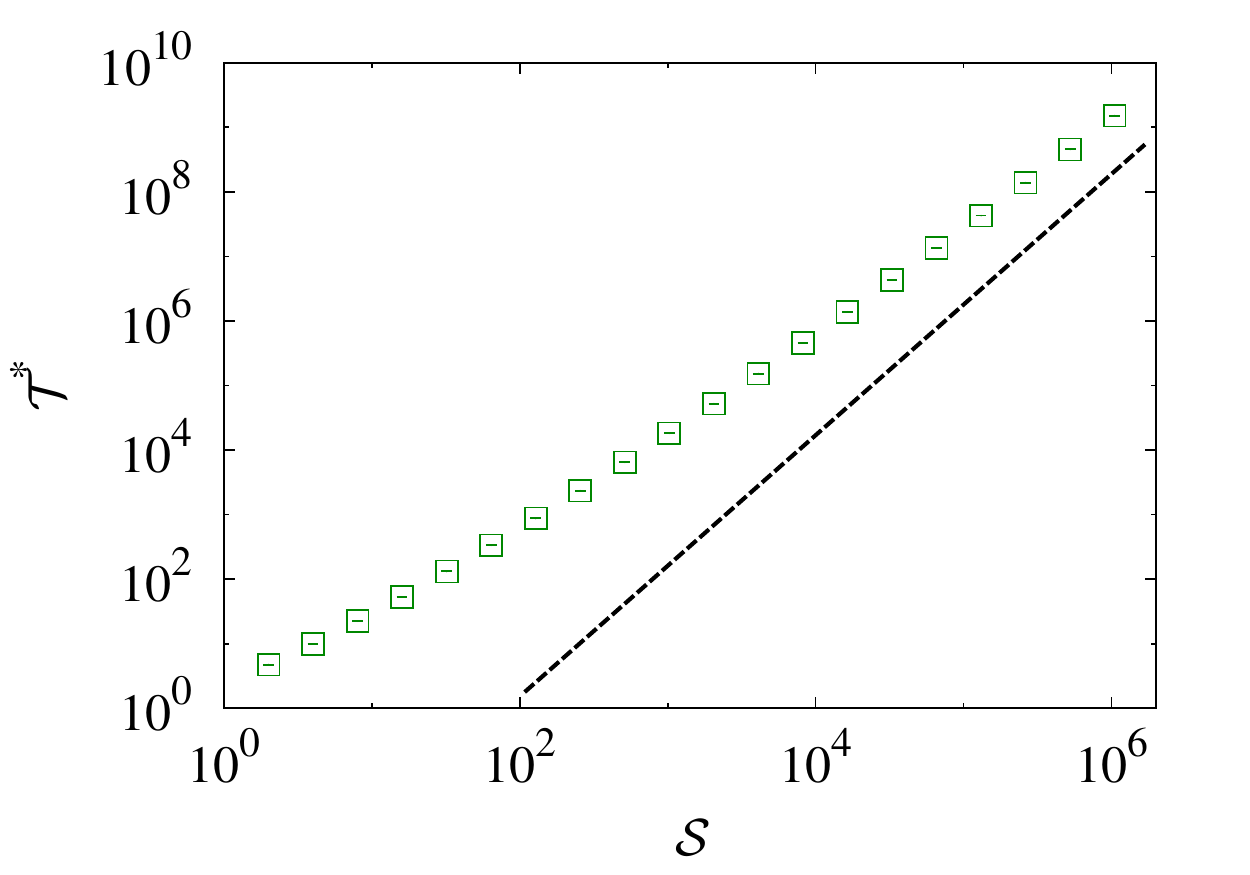}}}
\caption{(a) Optimal myopia $p^*$ versus $\mathcal{S}$, and (b) maximal
  lifetime versus $\mathcal{S}$ in $d=1$.  The (very thin) error bars are
  smaller than all data points.  The dashed lines have slope $-1$ in (a) and
  $+2$ in (b).  The data are based on $10^4$ realizations for
  $\mathcal{S}\leq 2^{19}$ and $10^3$ realizations for $\mathcal{S}=2^{20}$.}
\label{p-vs-S}
\end{figure}

We characterize the forager dynamics by its lifetime
$\mathcal{T(\mathcal{S}},p)$.  The basic feature of the myopic forager is
that there is an optimal value of the myopia parameter, $p^*(\mathcal{S})$,
distinct for each $\mathcal{S}$, that maximizes the forager lifetime
(Fig.~\ref{t-vs-p-1d}).  From plots such as these, we thereby determine the
optimal value of $p^*(\mathcal{S})$ for $\mathcal{S}$ between $2$ and
$2^{20}$ with an accuracy of 1\% or less.  As shown in Fig.~\ref{p-vs-S}(a),
the data for the optimal value $p^*(\mathcal{S})$ is roughly consistent with
$p^*\sim 1/\mathcal{S}$, but the data exhibit a slight but decreasing
downward curvature.  Indeed the functional form
$p^*\sim\ln\mathcal{S}/\mathcal{S}$ fits the data quite well.  Notice that
this form for $p^*$ also corresponds to the limit of validity of the
heuristic approach given in Sec.~\ref{sec:heu}.

Once we determine the optimal value $p^*$ for each $\mathcal{S}$, we then
study the $\mathcal{S}$ dependence of the lifetime at this optimal myopia
$p^*(\mathcal{S})$.  We define this maximal lifetime as
$\mathcal{T}^*(\mathcal{S})$.  The maximal lifetime is a smoothly increasing
function of $\mathcal{S}$ with slow upward curvature on a double logarithmic
scale (Fig.~\ref{p-vs-S}(b)).  This slow curvature again suggests the
presence of logarithmic corrections.  Indeed, the data for
$\ln\mathcal{S}\times \mathcal{T}^*$ appears to grow as $\mathcal{S}^2$.  In
fact, substituting the value of $p^*\sim \ln\mathcal{S}/\mathcal{S}$ into
Eq.~\eqref{T} also gives $\mathcal{T}^* \sim \mathcal{S}^2/\ln\mathcal{S}$.
Finally, notice that this optimal value of $\mathcal{T}^*$ is much larger
than the lifetime of the normal forager, which grows as $A\mathcal{S}$, with
$A$ exactly calculable and approximately equal to $3.2768$~\cite{BR14,CBR16}
and also much larger than the limiting $p\to 0$ behavior of
$\mathcal{S}=\mathcal{S}$.  Thus our heuristic argument predicts that there
must be maximum in the forager lifetime as a function of $p$, as well as the
$\mathcal{S}$ dependence of the maximal lifetime $\mathcal{T}^*$.

\begin{figure}[H]
  \center{\includegraphics[width=0.45\textwidth]{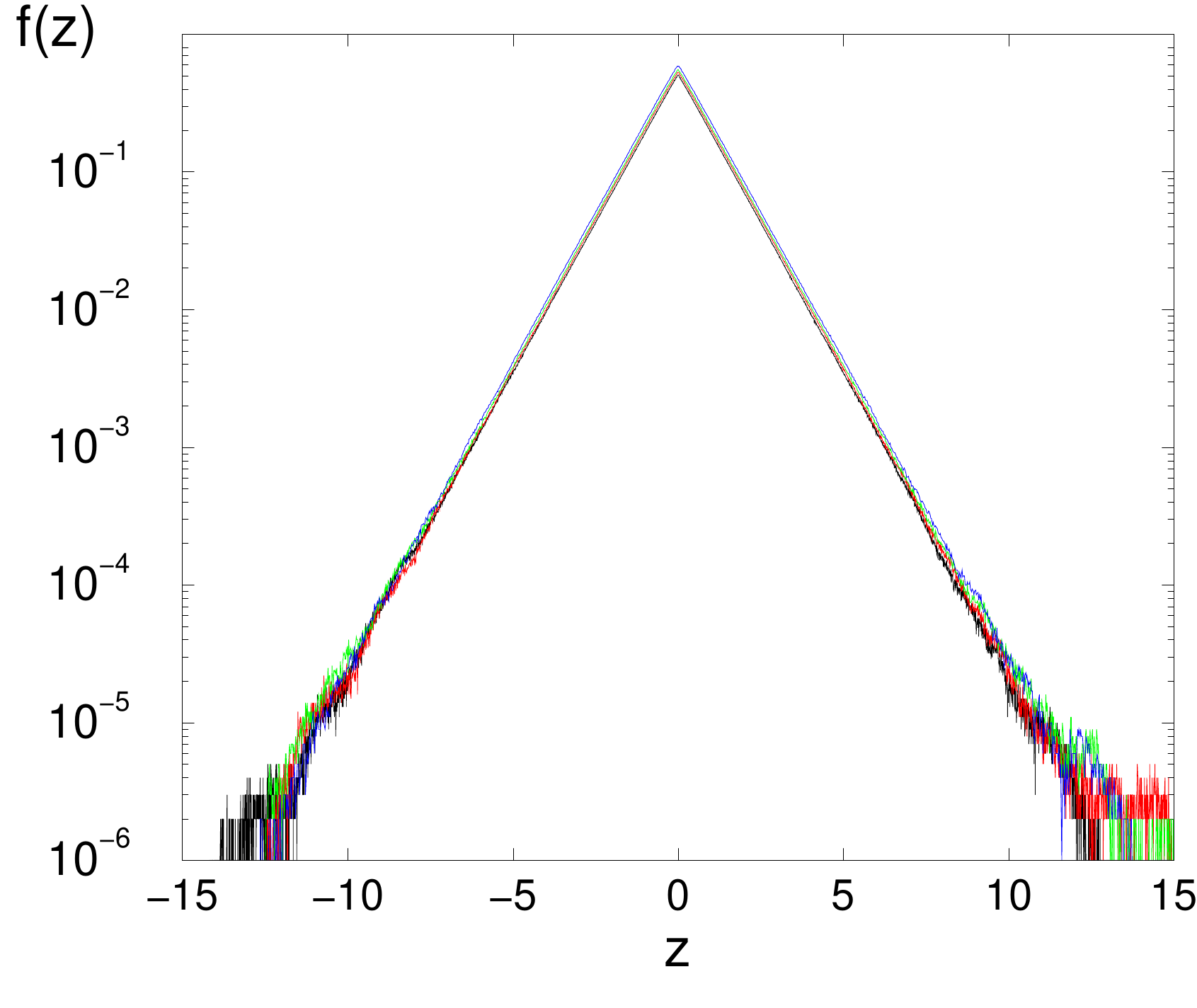}}
  \caption{ Scaled density profile $f(z)= X P(x/X)$ as a function of
    $z\equiv x/X$ in $d=1$.  The four datasets, each based on $10^6$
    realizations, are for $p=0.01$ and $\mathcal{S}=327= 10^{10/4}$ (black)
    $563=10^{11/4}$ (red), $1000=10^{12/4}$ (green), and $1779=10^{13/4}$
    (blue).}
  \label{profile}
\end{figure}

We now study the geometry of the desert.  Although the desert that is carved
out by the forager consists of multiple segments of empty and food-containing
sites (Fig.~\ref{space-time}), its average density profile has a simple
character (Fig.~\ref{profile}).  For a forager that starts at $x=0$ and has
metabolic capacity $\mathcal{S}$ and myopia parameter $p$, we measure the
probability $P(x)$ that food at site $x$ has been consumed up to the instant
when the forager starves.  The dependence of this probability distribution on
$p$ and $\mathcal{S}$ is not written for notational simplicity.  Clearly
$P(x)$ is decreases with $x$ because it is progressively less likely that the
forager reaches a large distance and consumes food there.  As shown in
Fig.~\ref{profile}, the density profile is an exponentially decaying function
of $x$ for all $x$; that is, $P(x)= X^{-1} \exp[-x/X]$, where
$X = \langle x(p,\mathcal{S})\rangle$ is the mean extent of the depleted
region.  Thus the scaling function $f(z)\equiv X\, P(x/X)$ depends only on
the scaling variable $z\equiv x/X$, as illustrated in the figure.

A simple mechanism underlies this exponential behavior of the density
profile.  Empirically, we find that the distribution of lifetimes
$Q(\mathcal{T})$ of the myopic forager has an exponential tail,
$Q(\mathcal{T})\sim \exp(-\mathcal{T}/\mathcal{T}^*)$ in all dimensions.  At
the instant of starvation the probability that the forager has traveled a
distance $r$ from its starting point is just the standard Gaussian
$p(x,\mathcal{T}) \sim \exp(-r^2/4D\mathcal{T})$, where $D$ is the diffusion
coefficient of the forager.  Convolving this Gaussian with the exponential
lifetime distribution $Q(\mathcal{T})$, the outcome is again exponential
decay in $x$, as written above.  Related convolution-generated non-Gaussian
behavior has been obtained in other generalized random walk
models~\cite{CSMS17}.

\section{Simulations in Greater Than One Dimension}

\begin{figure}[ht]
\center{\includegraphics[width=0.55\textwidth]{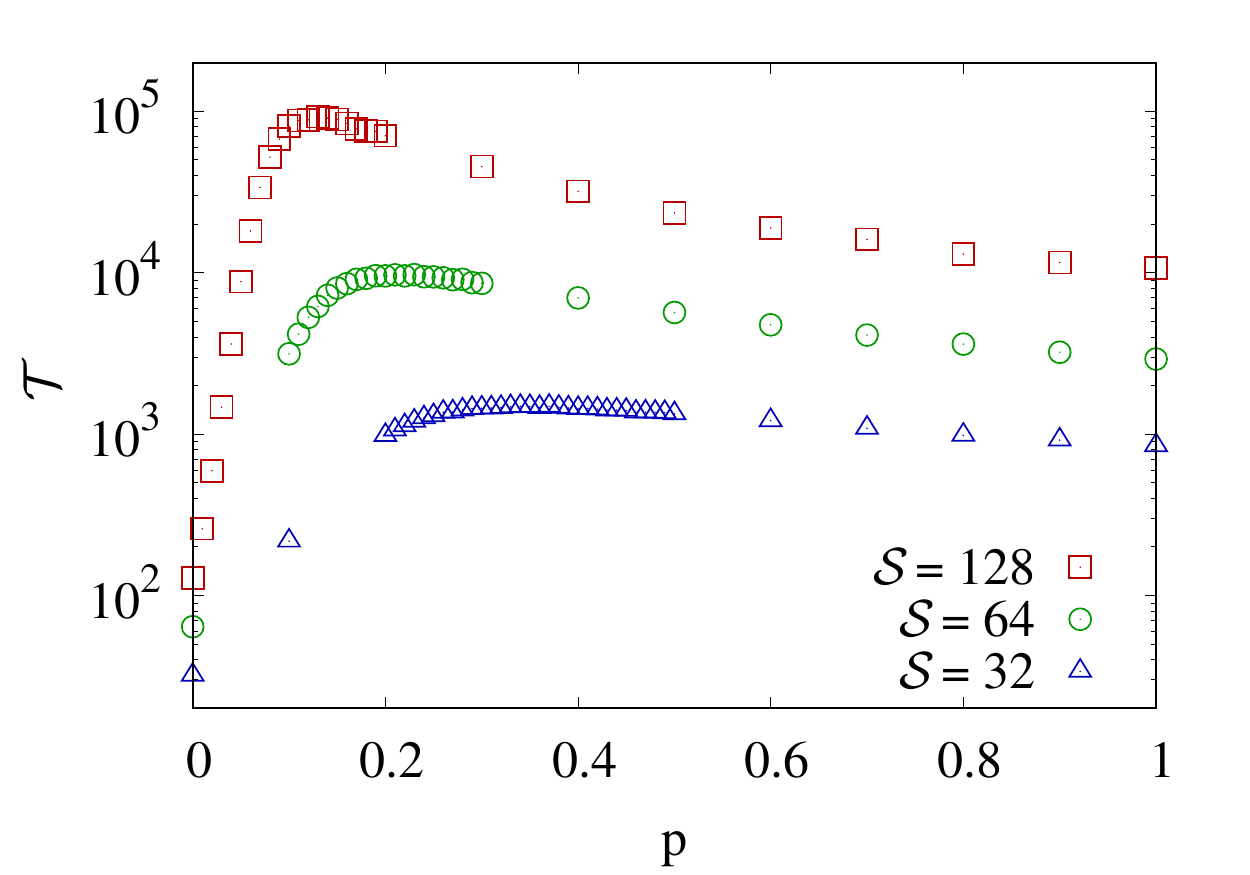}}
\caption{Forager lifetime versus $p$ in $d=2$ for three representative values
  of $\mathcal{S}$.}
  \label{t-vs-p-2d}
\end{figure}

\begin{figure}[ht]
\centerline{\subfigure[]{\includegraphics[width=0.45\textwidth]{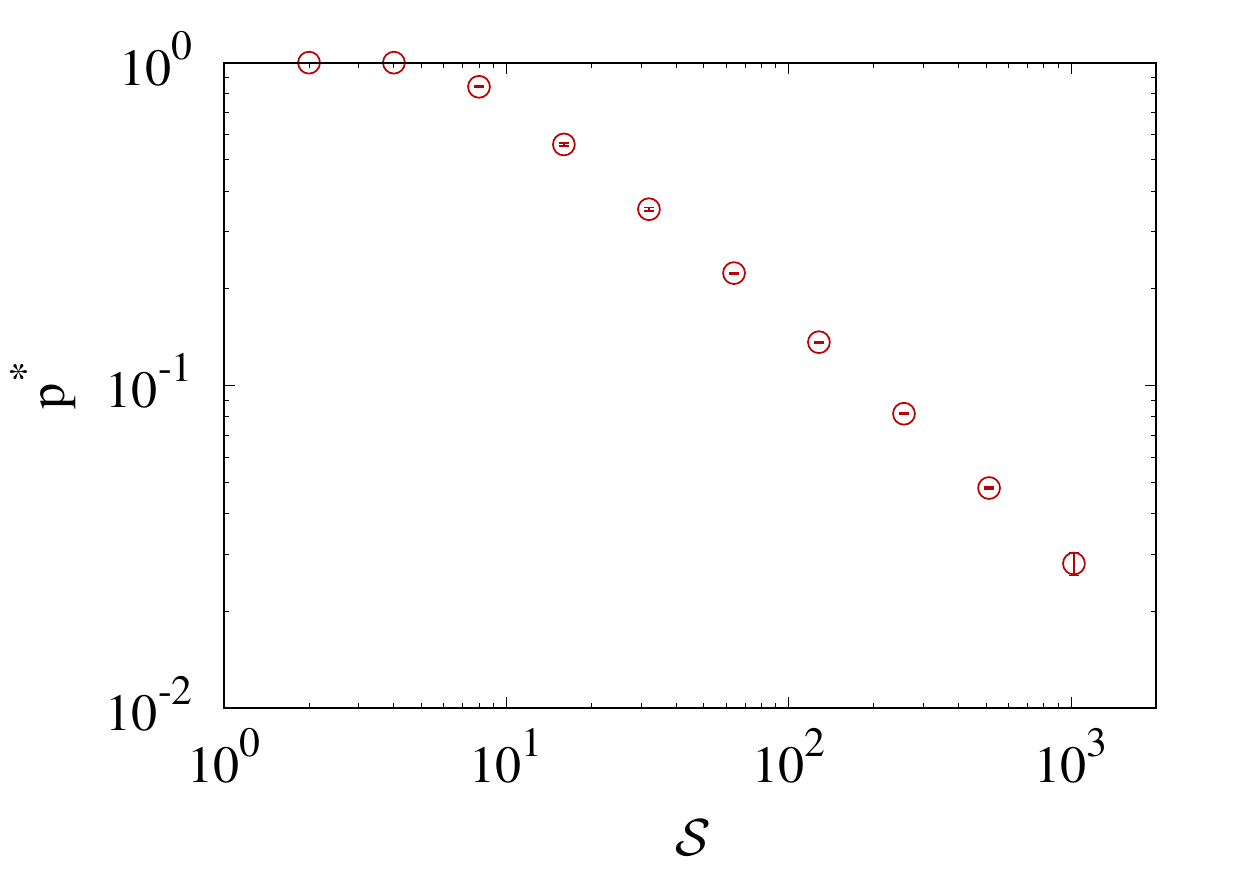}}
\subfigure[]{\includegraphics[width=0.45\textwidth]{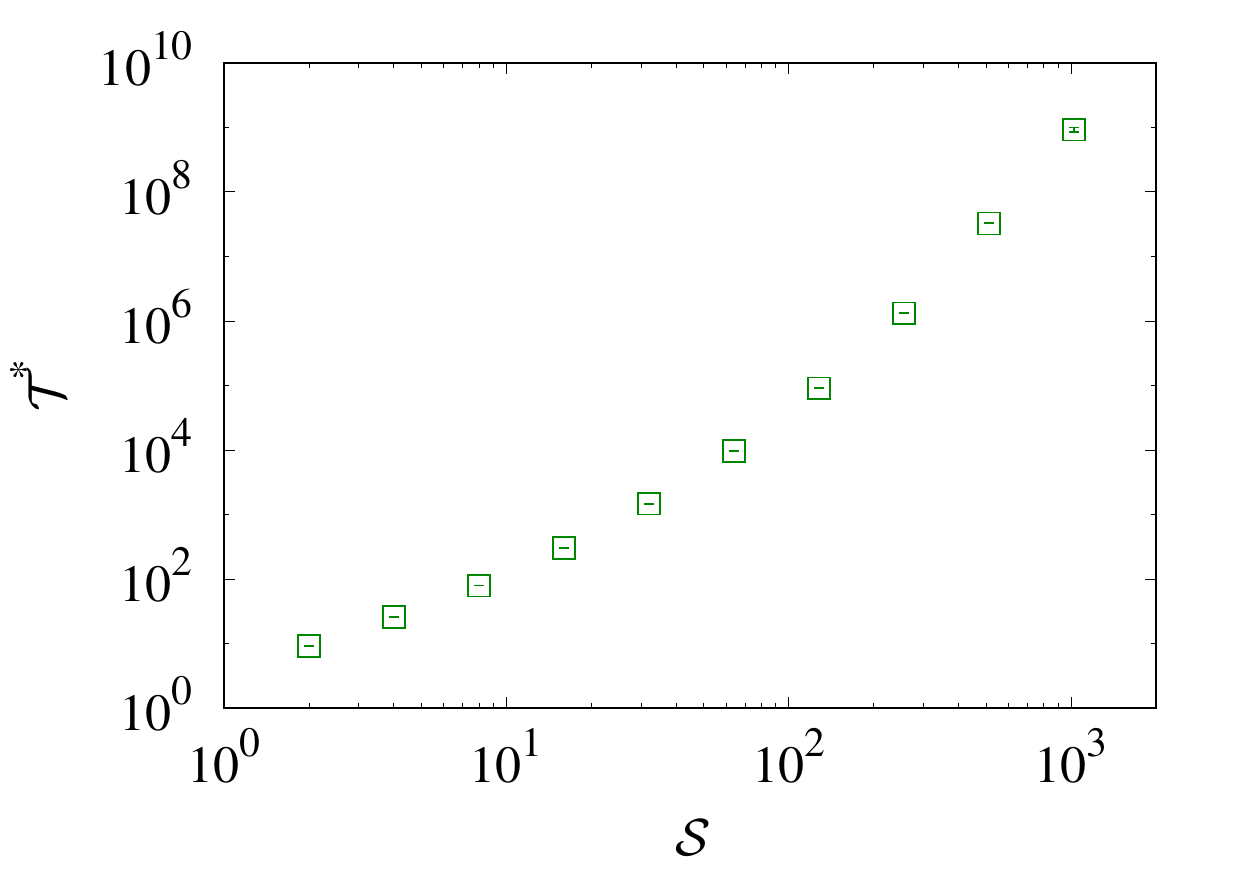}}}
\caption{(a) The optimal value of $p$, $p^*$, as a function of $\mathcal{S}$,
  and (b) the maximal forager lifetime, $\mathcal{T}^*$, as a function of
  $\mathcal{S}$ in $d=2$.  Error bars are shown.  The data is based
  on $10^4$ realizations for $\mathcal{S}\leq 2^9$ and $10^2$ realizations
  for $\mathcal{S}=2^{10}$. }
  \label{tpstar-2D}
\end{figure}

The dynamical behavior of the forager in $d\geq 2$ is qualitatively similar
to that in $d=1$.  However, because we can simulate only to
$\mathcal{S}=2^{10}$ in $d=2$ and to $\mathcal{S}=2^5$ in $d=3$, our
estimates for asymptotic behavior are imprecise.  In $d=2$, the forager
lifetime is again maximal at an intermediate value of $p^*$ that is strictly
between 0 and 1 (Fig.~\ref{t-vs-p-2d}) and also is a decreasing function of
$\mathcal{S}$ (Fig.~\ref{tpstar-2D}(a)).  The dependence of $p^*$ versus
$\mathcal{S}$ is almost linear on a double logarithmic scale (based on the
last 7 points).  A linear least-squares fit of the last 4 data points
indicates that $p^*\sim \mathcal{S}^{-\alpha}$, with $\alpha\approx 0.76$.
While the local slopes of $\ln p^*$ versus $\ln\mathcal{S}$ are becoming
slightly more negative for larger $\mathcal{S}$, the number of data points is
to few to extrapolate with any confidence.  Thus we believe that
$\alpha\approx 0.8$, with an uncertainty of roughly 0.1.

The maximal lifetime $\mathcal{T}^*$ of the myopic forager in $d=2$ grows
much more rapidly with $\mathcal{S}$ than in $d=1$.  On a double logarithmic
scale, the data show significant upward curvature, which suggests that
$\mathcal{T}^*$ grows faster than a power law in $\mathcal{S}$.  However, a
plot of $\ln \mathcal{T}^*$ versus $\mathcal{S}$ is curved downward, which
excludes exponential growth.  The data can be reasonably fit by a fractional
exponential, $\mathcal{T}\sim \exp(\mathcal{S}^\beta)$ with
$\beta\approx 0.3$.  This behavior roughly accords with Eq.~\eqref{T}: if
$p^*\sim \mathcal{S}^{-\alpha}$, with $\alpha\approx 0.8$, then Eq.~\eqref{T}
predicts that $\mathcal{T}\sim \exp(\mathcal{S}^{\beta})$, with
$\beta=1-\alpha =0.2$.  To reach an unambiguous conclusion about the
dependence of $\mathcal{T}^*$ versus $\mathcal{S}$ would require orders of
magnitude longer simulations.

\begin{figure}[ht]
\center{\includegraphics[width=0.45\textwidth]{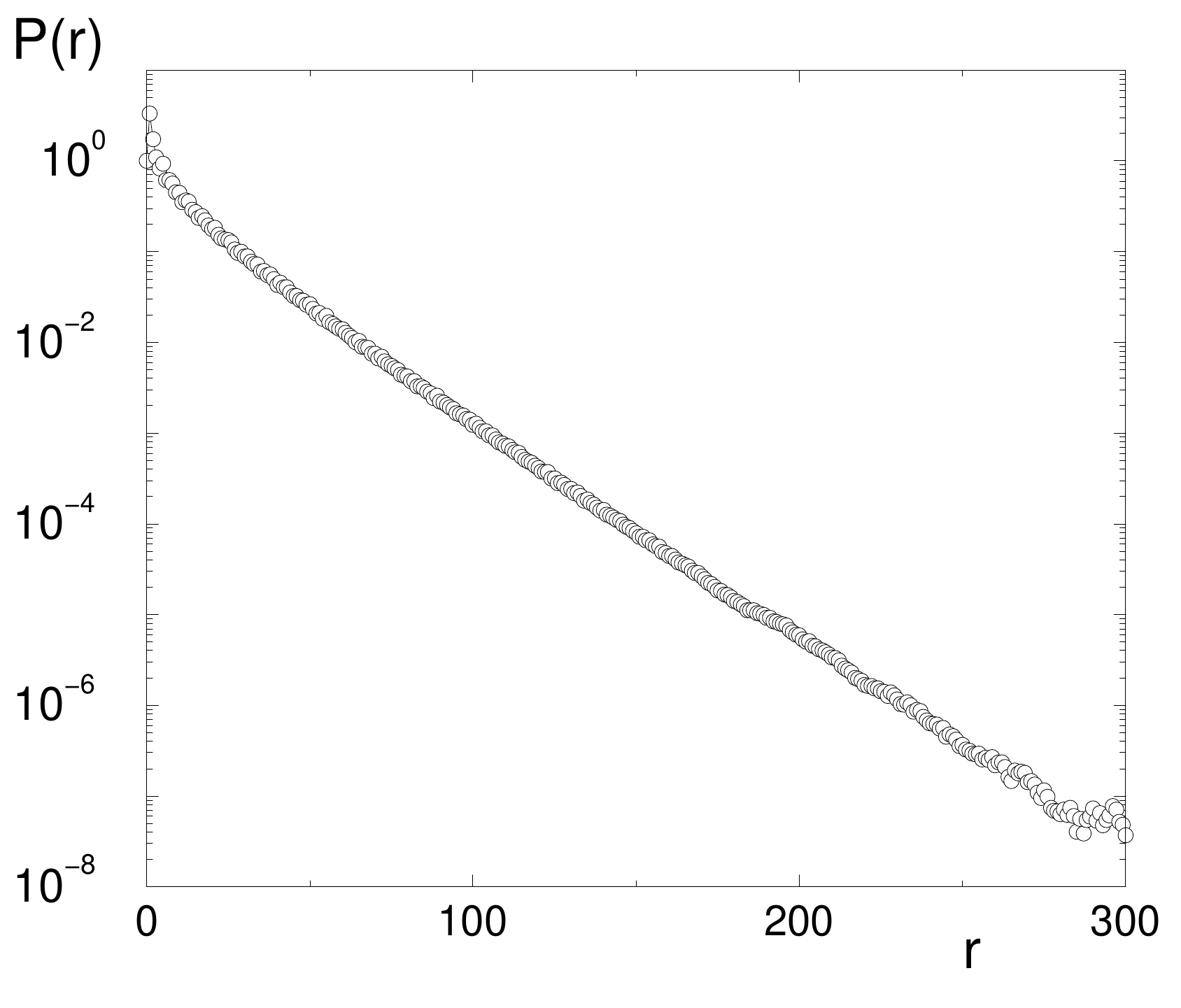}}
\caption{The radial density profile of the desert in $d=2$ for the case of
  $p=0.4$ and $\mathcal{S}=32$, averaged over $10^4$ realizations.}
  \label{radial-profile-2d}
\end{figure}

As in the case of $d=1$, the desert that is carved out by a forager in $d=2$
consists of multiple, disjoint food-free regions (Fig.~\ref{space-time}(d)).
In spite of this complicated geometry for a single trajectory, the average
profile of the desert again has a simple character.  For a forager that
starts at $x=0$ and has metabolic capacity $\mathcal{S}$ and myopia parameter
$p$, we measure the density profile $P(\mathbf{r})$ that the food at site
$\mathbf{r}$ has been consumed up to the instant that the forager starves.
This distribution is again a decreasing function of $|\mathbf{r}|$ because it
is progressively less likely that the forager reaches a large distance and
consumes food there (Fig.~\ref{radial-profile-2d}).  The data indicate that
the decay of the density profile is exponential in $\mathbf{r}$, as in the
case of one dimension.  The mechanism that causes this exponential density
profile is the same as that in one dimension.

\begin{figure}[ht]
  \centerline{\subfigure[]{\includegraphics[width=0.33\textwidth]{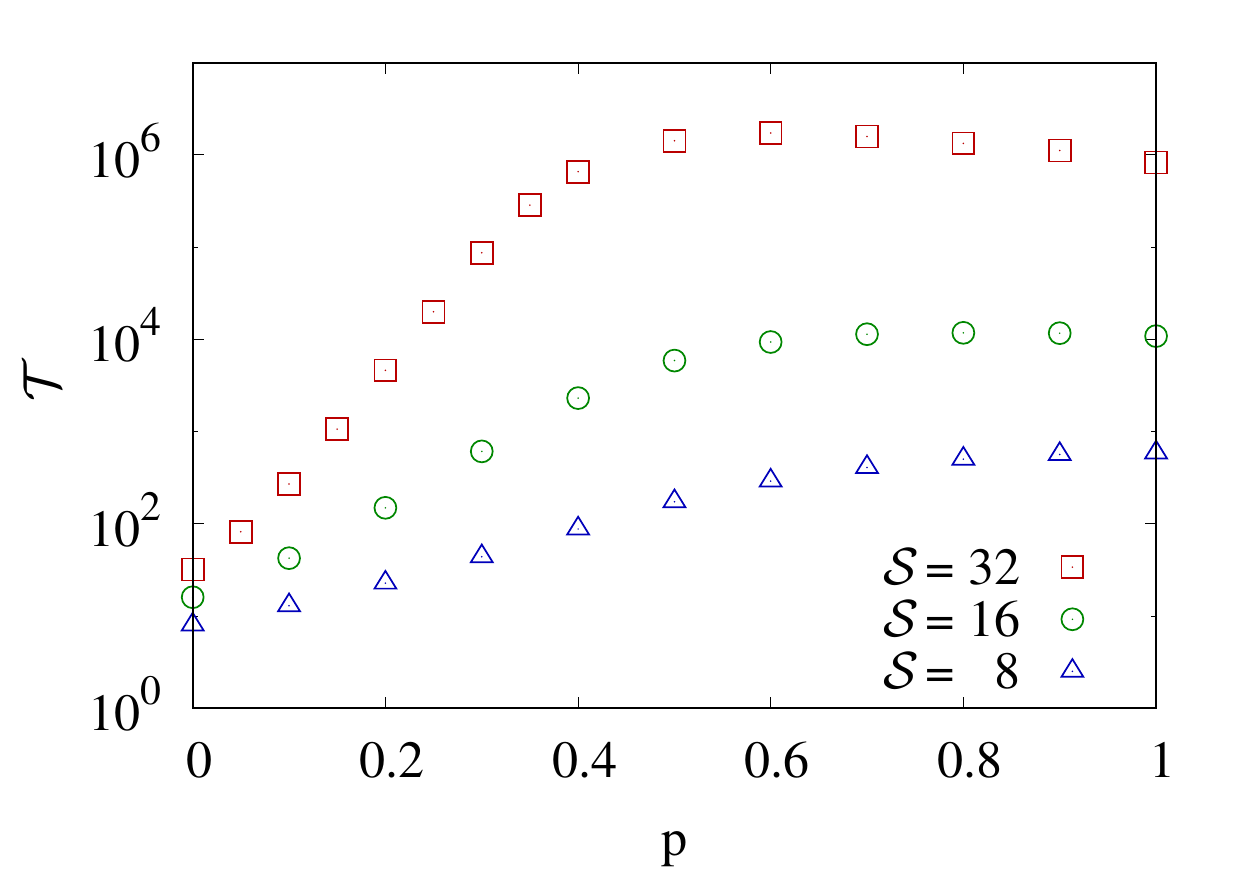}}
    \subfigure[]{\includegraphics[width=0.32\textwidth]{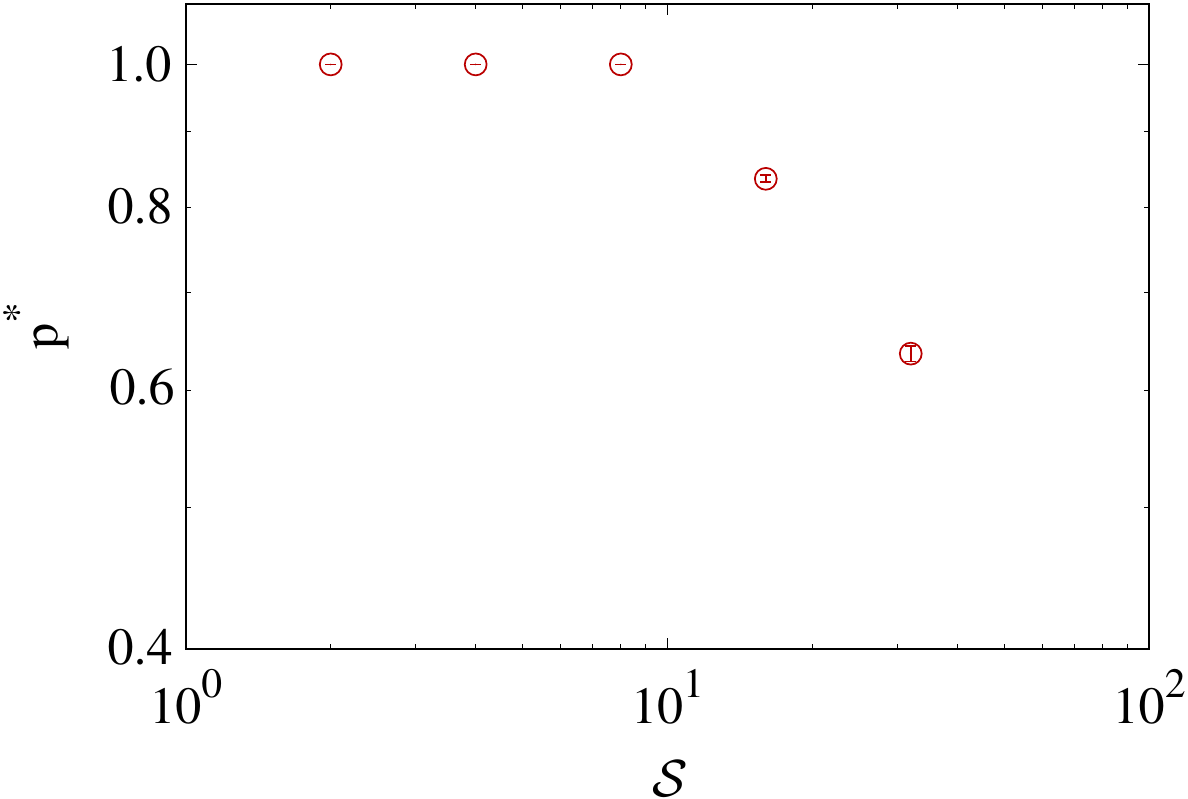}}
\subfigure[]{\includegraphics[width=0.33\textwidth]{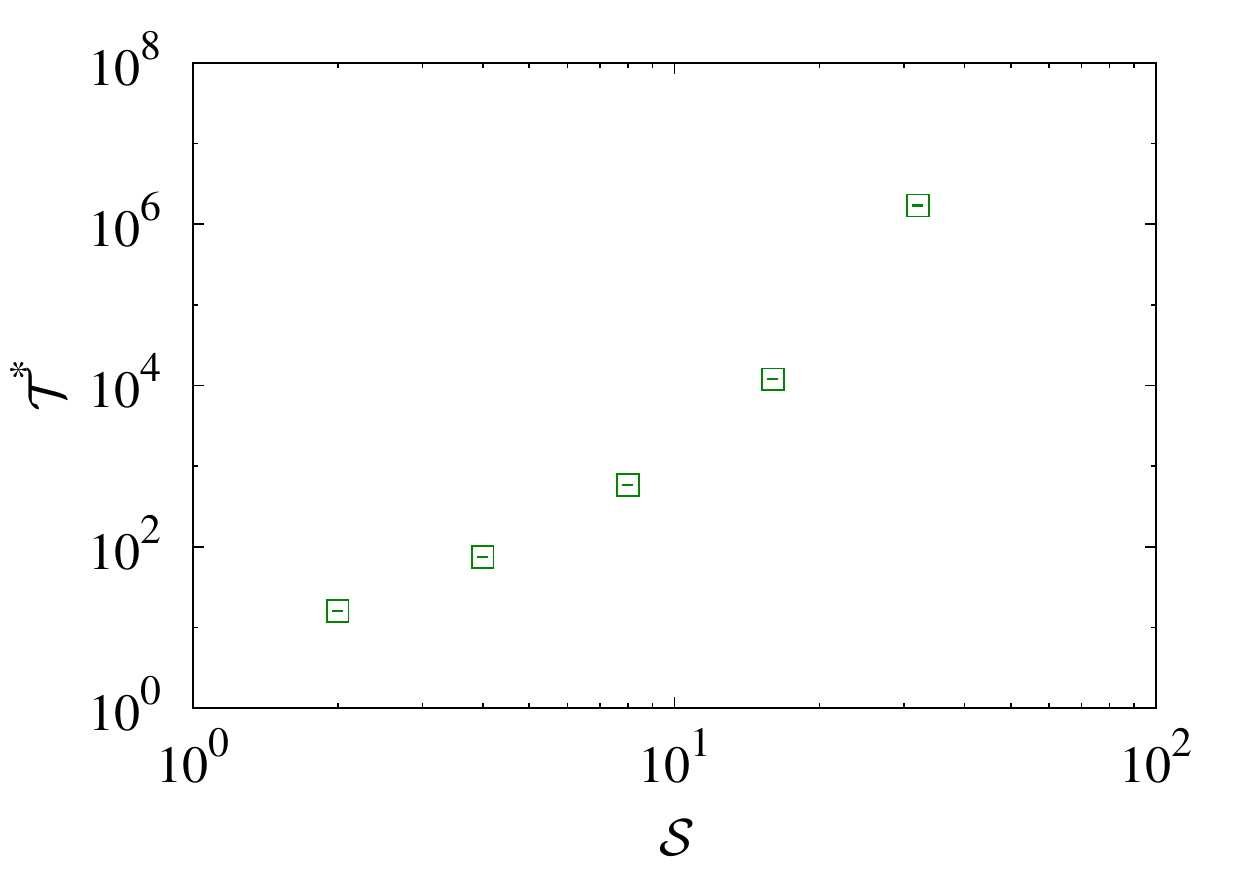}}}
\caption{Simulation results in $d=3$. (a) Forager lifetime versus $p$ for
  three representative $\mathcal{S}$ values.  (b) The optimal value of $p$,
  $p^*$, as a function of $\mathcal{S}$, and (c) the maximal forager
  lifetime, $\mathcal{T}^*$, as a function of $\mathcal{S}$.  Error bars are
  shown.  The data in (b) and (c) is based on $10^4$ realizations for
  $\mathcal{S}\leq 2^4$ and $10^3$ realizations for $\mathcal{S}=2^{5}$. }
  \label{3d-plots}
\end{figure}

For completeness, we also simulate the myopic forager in $d=3$.  Here the
maximal lifetime grows so rapidly with $\mathcal{S}$ and the requisite memory
needs are so large that we can only simulate the myopic forager for
$\mathcal{S}\leq 32$.  Figs.~\ref{3d-plots}(a) shows the behavior of the
lifetime versus $\mathcal{S}$ up to $\mathcal{S}=32$, while
Figures~\ref{3d-plots}(b) and (c), show the dependence of the optimal myopia
$p^*$ and the maximal lifetime $\mathcal{T}^*$ as a function of
$\mathcal{S}$.  The only claim that we can make from the small range of data
is that $\mathcal{T}^*$ grows faster than any power law in $\mathcal{S}$.

\section{Summary}

We extended the starving random walk model of foraging to the situation where
the forager is myopic and eats with probability $p<1$ when it encounters
food.  As found previously in a variety of idealized foraging
models~\cite{BRB17a,BRB17b,BBKR18}, the forager lifetime is maximized when
the basic model parameter, the degree of myopia $p$, is set to an optimal
value.  This optimal myopia $p^*$ appears to scale as
$\ln\mathcal{S}/\mathcal{S}$ in one dimension and as $\mathcal{S}^{-\alpha}$
in two dimensions with $\alpha \approx 0.8$.  At this optimal myopia, the
maximal lifetime appears to grow as $\mathcal{S}^2/\ln\mathcal{S}$ in one
dimension and as $\exp(\mathcal{S}^\beta)$ in two dimensions, with $\beta$
close to the value $1-\alpha$, as anticipated from Eq.~\eqref{T}.

An important message from this model is that a forager with a poor ability to
detect food lives much longer than a forager with perfect detection
capability.  This increased lifetime arises because the myopic forager
typically eats when it is nutritionally depleted by a significant amount, so
that the wastage of food resources is small.  In contrast, a normal forager
(with $p=1$) always eats when food is encountered and thus may waste a
considerable amount of food whenever it eats again soon after its most recent
meal.  Thus for a naive forager with limited information processing
capability, being myopic---equivalently, being somewhat clueless---turns out
to be a surprisingly effective survival strategy.

\section*{Acknowledgments}

We acknowledge support from the European Research Council Starting Grant No.\
FPTOpt-277998 (OB), a University of California Merced postdoctoral
fellowship (UB), and Grant No.\ DMR-1608211 from the National Science
Foundation (UB and SR).

\bigskip\bigskip%\newpage


\begin{thebibliography}{99}

\bibitem{BR14} O. B\'enichou and S. Redner, Phys.\ Rev.\ Lett.\ {\bf 113},
  238101 (2014).

\bibitem{CBR16} M. Chupeau, O. B\'enichou, and S. Redner, J. Phys.\ A: Math.\
  \& Theor.\ {\bf 49}, 394003 (2016).

\bibitem{MP66} R. H. MacArthur and E. R. Pianka, Am.\ Nat.\ \textbf{100},
603 (1966).

\bibitem{C76} E. L. Charnov, Theor.\ Popul.\ Biol.\ {\bf 9}, 129 (1976).

\bibitem{PPC77} G. H. Pyke, H. R. Pulliam, and E. L. Charnov, Q. Rev.\ Biol.\ \textbf{52},
  137 (1977).
  
\bibitem{SK86} D. W. Stephens and J. R. Krebs, {\it Foraging Theory},
  (Princeton University Press, Princeton, NJ, 1986).

\bibitem{OB90} W. J. O'Brien, H. I. Browman, and B. I. Evans, Am.\ Sci.\ {\bf
    78}, 152 (1990).

\bibitem{B91} J. W. Bell, Searching Behaviour, the Behavioural Ecology of
  Finding Resources, Animal Behaviour Series (Chapman and Hall, London,
  1991).

\bibitem{BRB17a} U. Bhat, S. Redner, and O. B\'enichou, Phys.\ Rev.\ E {\bf
    95}, 062119 (2017).

\bibitem{BRB17b} U. Bhat, S. Redner, and O. B\'enichou, J. Stat.\ Mech.\ 073213
  (2017).

\bibitem{BBKR18} O. B\'enichou, U. Bhat, P. L. Krapivsky, and S. Redner,
  Phys.\ Rev.\ E \textbf{97}, 022110 (2018).

\bibitem{H74} J. J. Hopfield, Proc.\ Nat.\ Acad.\ Sci.\ (USA) \textbf{71} 4135
(1974).

\bibitem{MBN09} B. Munsky, G. Bel, and I. Nemenman, J. Chem.\ Phys.\
\textbf{131}, 235103 (2009).

\bibitem{BMN10} G. Bel, B. Munsky, and I. Nemenman, Phys.\ Biol.\
\textbf{7}, 016003 (2010).
  
\bibitem{R01} S. Redner, {\it A Guide to First-Passage Processes}, (Cambridge
  University Press, Cambridge, UK, 2001).

\bibitem{CSMS17} A. V. Chechkin, F. Seno, R. Metzler, I. M. Sokolov, Phys.\
  Rev.\ X \textbf{7}, 021002 (2017).
  
\end{thebibliography}
\end{document}